\documentclass[prd,onecolumn,nofootinbib,aps,superscriptaddress,tightenlines,preprintnumbers]{revtex4}

\usepackage{epsfig}
\usepackage{amsmath}

  \newcount\hour \newcount\minute
  \hour=\time \divide \hour by 60
  \minute=\time
  \newcommand{\mydate}{\ \today \ - \number\hour :\ifnum \minute<10 0\fi
\number\minute}

\def\nn{\nonumber \\ }

\newcommand{\pbra}[1]{\left(#1\right)}
\newcommand{\bbra}[1]{\left[#1\right]}

\begin{document}

\preprint{\hbox{CALT-68-2889}  }

\title{Decoherence Problem in Ekpyrotic Phase}

\author{Chien-Yao Tseng}
\affiliation{California Institute of Technology, Pasadena, CA 91125 }

\begin{abstract}
Quantum decoherence  and the transition to semiclassical behavior during inflation has been extensively considered in the literature. In this paper, we use a simple model to analyze the same process in ekpyrosis. Our result is that the quantum to classical transition would not happen during an ekpyrotic phase even for superhorizon modes, and therefore the fluctuations cannot be interpreted as classical. This implies the prediction of scale-free power spectrum in ekpyrotic/cyclic universe model requires more inspection. 
\end{abstract}

\maketitle
\section{Introduction}
From cosmological observations we know that the current universe is to a good approximation flat, homogeneous and isotropic on large scales \cite{COBE,WMAP3}. It is well known that in standard Big Bang cosmology this requires an enormous amount of fine-tuning on the initial conditions. Two mechanisms are provided to be possible explanations. The first is inflation \cite{Guth:1980zm,Linde:Albrecht}, a period of accelerated expansion occuring between the big bang and nucleosynthesis. The second is ekpyrosis \cite{Khoury:2001wf,Lehners:2008vx,Erickson:2003zm,Khoury:2001bz,Turok:2004gb}, a period of ultra-slow contraction before Big Bang/Big Crunch to an expanding phase. Both mechanisms not only manage to address the standard cosmological puzzles but also have the ability to imprint scale invariant inhomogeneities on superhorizon scales via a causal mechanism \cite{Guth:1980zm,Khoury:2001wf,Mukhanov:1981xt,Hawking:1982cz,Starobinsky:1982ee,Bardeen:1983qw,Erickson:2006wc}. These inhomogeneities are thought to provide the seeds which later become the temperature anisotropies in the Cosmic Microwave Background and the Large Scale Structure in the universe. This framework of the cosmological perturbation theory is based on the quantum mechanics of scalar fields, where the relevant observable is the amplitude of the field's Fourier modes \cite{inflationreview}. Although treated as a quantum mechanical variable, this amplitude is expected to be stochastic variables, characterized by averages of their products, i.e., power spectrum. This interpretation proves to be very accurate in the CMB and Large Scale structure analyses.

However, in order to make this stochastic interpretation consistent, the density matrix has to be diagonal in the amplitude basis. This criterion implies that interference terms in the density matrix are highly suppressed and can be neglected \cite{Halliwell:1989vw,Padmanabhan:1989rm}. Interference is associated with the coherence of the system, i.e., the coherence in the state between different points of configuration space \cite{Zurek:1981xq,Joos:1984uk}. A measure of this is the coherence length which gives the configuration distance over which off-diagonal terms are correlated \cite{Brandenberger:1990bx}.

An isolated system described by the Schr\"{o}dinger equation cannot lose its coherence; a pure state always remains pure. However, if it is coarse grained, it may evolve from a pure to a mixed state. One way to realize coarse graining is to let the system interact with an environment \cite{Zurek:1981xq}. The environment consists of all fields whose evolution we are not interested in. The state of the system is obtained by tracing over all possible states of the environment. Now, even if the state describing system plus environment is pure, the state of the system alone will in general be mixed.

In the literature, there are various arguments and calculations suggesting that a form of such environment decoherence can indeed occur for inflationary perturbations \cite{Brandenberger:1990bx,Calzetta:1995ys,Lombardo:1995fg,Matacz:1996gk,Campo:2005sv,Martineau:2006ki,Burgess:2006jn,Kiefer:2006je,Lombardo:2005iz,Prokopec:2006fc,Koksma:2007zz}. The coherence length decreases exponentially for wavelengths greater than Hubble radius. Thus perturbations become classical once their wavelength exceeds the Hubble radius. All of these results lend support to the usual heuristic derivation of the spectrum of density perturbations in inflationary models. In this paper, we use a simple model to study whether decoherence can also occur in the ekyprotic phase. We find that the coherence lengths continue increasing even for the modes outside the horizon.  Therefore, the heuristic argument that the modes become classical when they leave the horizon is invalid in the ekyprotic phase and requires more careful inspection.

\section{The model}
A crucial question is how to model the environment. Any realistic model will be very complicated and hard to analyze. However, the basic physics should emerge from the simplest models. Hence, we choose a model \cite{Brandenberger:1990bx} which can be solved exactly: the system is a real massless scalar field $\phi_1$, and the environment is taken to be a second massless real scalar field $\phi_2$ interacting with $\phi_1$ through their gradients.

The action of system and environment is
\begin{equation}
 S=\int d^4 x \mathcal{L}=\int d^4 x \sqrt{-g}~\frac 12 \pbra{-\partial_\mu\phi_1\partial^\mu\phi_1-\partial_\mu\phi_2\partial^\mu\phi_2-2c\partial_\mu\phi_1\partial^\mu\phi_2}
\end{equation}
where $g$ is the determinant of the background metric which is given by
\begin{equation}
 ds^2=a^2(\eta)(-d\eta^2+d\mathbf x^2)
\end{equation}
and $c\ll 1$ is the coupling constant describing the interaction between two fields. Note that this Lagrangian is quadratic in the derivative of the fields and can hence be diagonalized for which the interaction term disappears and the whole Lagrangian becomes a free field theory. If there is no other field or interaction in our universe, this argument is true. However, we suppose there is a hidden interaction such that we can only obeserve the first field $\phi_1$ but not the environment $\phi_2$. In other words, we assume the environment and the observed system do not form the diagonal basis. This assumption is reasonable since any observed scalar fields (whose reduced density matrix we want) will interact with gravitational perturbations (which is a part of the environment). 

Then, the canonical momenta $\pi_i$ conjugate to the fields $\phi_i$, $i=1,2$ are
\begin{eqnarray}
 \pi_1 &=& \frac{\partial \mathcal L}{\partial\dot\phi_1}=a^2\pbra{\dot\phi_1+c\dot\phi_2} \\
 \pi_2 &=& \frac{\partial \mathcal L}{\partial\dot\phi_2}=a^2\pbra{\dot\phi_2+c\dot\phi_1}
\end{eqnarray}
where ``$\cdot$'' denotes derivative with respect to $\eta$. This allows us to write the Hamiltonian $H$ as
\begin{equation}
 H=\int d^3x (\pi_i\dot\phi_i-\mathcal L)=\int d^3x\left\{\frac{1}{2a^2(1-c^2)}\pbra{\pi_1^2+\pi_2^2-2c\pi_1\pi_2}+\frac{a^2}{2}\bbra{(\mathbf\nabla\phi_1)^2+(\mathbf\nabla\phi_2)^2+2c(\mathbf\nabla\phi_1)\cdot(\mathbf\nabla\phi_2)}\right\}
\end{equation}
To study decoherence, it is more convenient to use the functional Schr\"odinger picture\cite{Hill:1985gw}. The commutation relation $\bbra{\phi_i(\mathbf x),\pi_j(\mathbf{y})}=i\delta_{ij}\delta^3(\mathbf x-\mathbf y)$ is equivalent to making the replacement $\pi_i(\mathbf x)\rightarrow -i\frac{\delta}{\delta\phi_i(\mathbf x)}$ . The wave functional $\Psi[\phi_1,\phi_2]$ obeys the Schr\"odinger equation
\begin{equation}
\label{schrodinger}
 i\frac{\partial}{\partial\eta}\Psi=\hat H\Psi
\end{equation}
We make a Gaussian ansatz for $\Psi$ to be able to find the vacuum or ground state solution:
\begin{equation}
\label{wave}
 \Psi[\phi_1,\phi_2]=\mathcal N\exp\bbra{-\frac 12\int d^3x d^3y \pbra{\phi_1(\mathbf x)\phi_1(\mathbf y)+\phi_2(\mathbf x)\phi_2(\mathbf y)}A(\mathbf x,\mathbf y, \eta)+2\phi_1(\mathbf x)\phi_2(\mathbf y)B(\mathbf x,\mathbf y, \eta)}
\end{equation}
Note that we have already used the $\phi_1\leftrightarrow\phi_2$ symmetry of the Lagrangian. Furthermore, because of the $x \leftrightarrow y$ symmetry of the above integration, we have to require
\begin{eqnarray}
\label{symmetry}
 A(\mathbf x, \mathbf y,\eta) &=& A(\mathbf y, \mathbf x,\eta)\\
 B(\mathbf x, \mathbf y,\eta) &=& B(\mathbf y, \mathbf x,\eta)
\end{eqnarray}
Plug Eq.~\eqref{wave} into Schr\"odinger equation \eqref{schrodinger}, it is not difficult to get
\begin{eqnarray}
\label{eq1}
 \frac i2\frac{\partial A(\mathbf x, \mathbf y, \eta)}{\partial \eta} = \int d^3z \frac{1}{2a^2(1-c^2)}\bbra{A(\mathbf x,\mathbf z, \eta)A(\mathbf y,\mathbf z, \eta)+B(\mathbf x,\mathbf z, \eta)B(\mathbf y,\mathbf z, \eta)-2cA(\mathbf x,\mathbf z, \eta)B(\mathbf y,\mathbf z, \eta)}\nn
+\frac{a^2}{2}\mathbf\nabla^2_y\delta^3(\mathbf x-\mathbf y)
\end{eqnarray}
\begin{eqnarray}
\label{eq2}
 \frac i2\frac{\partial A(\mathbf x, \mathbf y, \eta)}{\partial \eta} = \int d^3z ~\frac{1}{2a^2(1-c^2)}\bbra{B(\mathbf x,\mathbf z, \eta)B(\mathbf y,\mathbf z, \eta)+A(\mathbf x,\mathbf z, \eta)A(\mathbf y,\mathbf z, \eta)-2cB(\mathbf x,\mathbf z, \eta)A(\mathbf y,\mathbf z, \eta)}\nn
+\frac{a^2}{2}\mathbf\nabla^2_y\delta^3(\mathbf x-\mathbf y)
\end{eqnarray}
\begin{eqnarray}
\label{eq3}
 \frac i2\frac{\partial B(\mathbf x, \mathbf y, \eta)}{\partial \eta} = \int d^3z ~\frac{1}{2a^2(1-c^2)}[2A(\mathbf x,\mathbf z, \eta)B(\mathbf y,\mathbf z, \eta)+2B(\mathbf x,\mathbf z, \eta)A(\mathbf y,\mathbf z, \eta)-2cA(\mathbf x,\mathbf z, \eta)A(\mathbf y,\mathbf z, \eta)\nn
-2cB(\mathbf x,\mathbf z, \eta)B(\mathbf y,\mathbf z, \eta)]+\frac{a^2}{2}\cdot 2c\mathbf\nabla^2_y\delta^3(\mathbf x-\mathbf y)
\end{eqnarray}
\begin{equation}
 i\frac{\partial \ln \mathcal N}{\partial \eta}=\frac{1}{2a^2(1-c^2)}\int d^3z \bbra{2A(\mathbf z,\mathbf z, \eta)-2B(\mathbf z,\mathbf z, \eta)}
\end{equation}

All the above equations come from the comparison of the coefficients in front of $\phi_i(\mathbf x)\phi_j(\mathbf y)$. It is easy to see that Eq.~\eqref{eq2} and Eq.~\eqref{eq1} are equivalent, which is just the result of the symmetry of $\phi_1$ and $\phi_2$. In order to satisfy Eq.~\eqref{eq1}-\eqref{eq3}, we have to require $B(\mathbf x,\mathbf y, \eta)=cA(\mathbf x,\mathbf y, \eta)$, which gives
\begin{equation}
 \Psi[\phi_1,\phi_2]=\mathcal N\exp\left\{-\frac 12\int d^3x d^3y\bbra{\phi_1(\mathbf x)\phi_1(\mathbf y)+\phi_2(\mathbf x)\phi_2(\mathbf y)+2c\phi_1(\mathbf x)\phi_2(\mathbf y)}A(\mathbf x,\mathbf y,\eta)\right\}
\end{equation}

\begin{eqnarray}
 i\frac{\partial \ln \mathcal N}{\partial \eta}&=&\frac{1}{a^2}\int d^3z A(\mathbf z,\mathbf z, \eta) \\
 i\frac{\partial A(\mathbf x, \mathbf y, \eta)}{\partial \eta}&=&\frac{1}{a^2}\int d^3z A(\mathbf x,\mathbf z, \eta)A(\mathbf y,\mathbf z, \eta)+a^2\mathbf\nabla^2_y\delta^3(\mathbf x-\mathbf y)
\label{position}
\end{eqnarray}
It is more convenient to solve Eq.~\eqref{position} in momentum space. Upon writing
\begin{eqnarray}
 \phi_i(\mathbf x) &=& \int \frac{d^3k}{(2\pi)^3} \phi_i(\mathbf k)e^{i\mathbf k\cdot\mathbf x}\\
 A(\mathbf x,\mathbf y, \eta) &=& \int \frac{d^3k}{(2\pi)^3} A(\mathbf k, \eta)e^{i\mathbf k\cdot(\mathbf x-\mathbf y)}
\end{eqnarray}
we get
\begin{equation}
\label{momentum}
 i\frac{\partial A(\mathbf k,\eta)}{\partial \eta}=\frac{1}{a^2}A^2(\mathbf k,\eta)-a^2k^2
\end{equation}
Here we have already used the relation $A(-\mathbf k,\eta)=A(\mathbf k,\eta)$ coming from Eq.~\eqref{symmetry}. Note that $A(\mathbf k, \eta)$ is only a function of $|\mathbf k|$, so we will write it as $A_k(\eta)$ from now on. This differential equation can be easily solved by assuming
\begin{equation}
\label{Ak}
 A_k(\eta)=-ia^2(\eta)\bbra{\frac{\dot u_k(\eta)}{u_k(\eta)}-\frac{\dot a(\eta)}{a(\eta)}}
\end{equation}
Then Eq.~\eqref{momentum} becomes
\begin{equation}
\label{uk}
 \ddot u_k+\pbra{k^2-\frac{\ddot a}{a}}u_k=0
\end{equation}
The wave functional can also be expressed in momentum space,
\begin{eqnarray}
 \Psi[\phi_1,\phi_2]&=&\mathcal N \exp\left\{-\frac 12\int \frac{d^3k}{(2\pi)^3} \bbra{\phi_1^*(\mathbf k)\phi_1(\mathbf k)+\phi_2^*(\mathbf k)\phi_2(\mathbf k)+c\phi_1^*(\mathbf k)\phi_2(\mathbf k)+c\phi_2^*(\mathbf k)\phi_1(\mathbf k)}A_k(\eta)\right\}\nn
&\equiv&\prod_k \Psi_k
\end{eqnarray}
where
\begin{equation}
\label{wavefn}
 \Psi_k=\mathcal N_k\exp\left\{-\frac 12\bbra{\phi_1^*(\mathbf k)\phi_1(\mathbf k)+\phi_2^*(\mathbf k)\phi_2(\mathbf k)+c\phi_1^*(\mathbf k)\phi_2(\mathbf k)+c\phi_2^*(\mathbf k)\phi_1(\mathbf k)}A_k(\eta)\right\}
\end{equation}
and $\phi_i(-\mathbf k)=\phi_i^*(\mathbf k)$ for the real scalar field. Because there is no coupling between modes with different $\mathbf k$, we will only consider a single wavelength and drop the index $\mathbf k$ for convenience from now on.

\section{The density matrix and the coherence length}
We now have the wave functional for all modes with single wavelength $\mathbf k$. The next step is to calculate the reduced density matrix for $\phi_1$ by tracing out $\phi_2$.
\begin{eqnarray}
 &&\rho(\phi_1,\bar\phi_1;\eta) = \int d\phi_2 d\phi^*_2~ \Psi^*_k(\phi_1,\phi_2,\eta)\Psi_k(\bar\phi_1,\phi_2,\eta)\\
 &&= |\mathcal N_k|^2\int d\phi_2 d\phi^*_2\exp\bbra{-\frac 12(\phi_1\phi^*_1+\phi_2\phi^*_2+c\phi_1\phi^*_2+c\phi_2\phi^*_1)A^*-\frac 12(\bar\phi_1\bar\phi^*_1+\phi_2\phi^*_2+c\bar\phi_1\phi^*_2+c\phi_2\bar\phi^*_1)A}
\end{eqnarray}
This can be computed from the Gaussian integral:
\begin{equation}
\label{reducedDM}
 \rho(\phi_1,\bar\phi_1;\eta) =\frac{4\pi}{A+A^*}|\mathcal N_k|^2 \exp\pbra{R+iI},
\end{equation}
where
\begin{eqnarray}
 R=-\frac{A+A^*}{4}\pbra{|\phi_1|^2+|\bar\phi_1|^2}+\frac{c^2}{8(A+A^*)}[(A+A^*)^2[\pbra{|\phi_1|^2+|\bar\phi_1|^2+\phi^*_1\bar\phi_1+\phi_1\bar\phi^*_1}\nn
   +(A^*-A)^2\pbra{|\phi_1|^2+|\bar\phi_1|^2-\phi^*_1\bar\phi_1-\phi_1\bar\phi^*_1}]
\end{eqnarray}
\begin{equation}
 iI=-(1-c^2)\frac{A^*-A}{4}\pbra{|\phi_1|^2-|\bar\phi_1|^2}
\end{equation}
To determine the coherence length of the reduced density matrix, it is convenient to introduce the new variables:
\begin{eqnarray}
 \chi&\equiv&\frac 12(\phi_1+\bar\phi_1)\\
 \Delta&\equiv&\frac 12(\phi_1-\bar\phi_1)
\end{eqnarray}
In terms of these variables, the reduced density matrix \eqref{reducedDM} becomes
\begin{equation}
 \rho(\phi_1,\bar\phi_1;\eta)=\frac{4\pi}{A+A^*}|\mathcal N_k|^2\exp\bbra{-\pbra{\frac{|\chi|^2}{\sigma^2}+\frac{|\Delta|^2}{l_c^2}+\beta(\chi\Delta^*+\chi^*\Delta)}}
\end{equation}

Because $\beta=\frac{1-c^2}{2}(A^*-A)$ is purely imaginary, the third term in the exponential just gives a complex phase. The first term gives the dispersion of the system, the dispersion coefficient $\sigma$ being
\begin{equation}
 \sigma=\sqrt{\frac{2}{(1-c^2)(A+A^*)}}
\end{equation}
The second term describes how fast the density matrix decays when considering the off-diagonal terms. Hence, $l_c$ is called the coherence length and is given by
\begin{equation}
\label{lc}
 l_c=\sqrt{\frac{2}{(A+A^*)\bbra{1-c^2\pbra{\frac{A^*-A}{A+A^*}}^2}}}
\end{equation}

\section{Decoherence in the usual inflation model}
\begin{figure}[ht]
\includegraphics{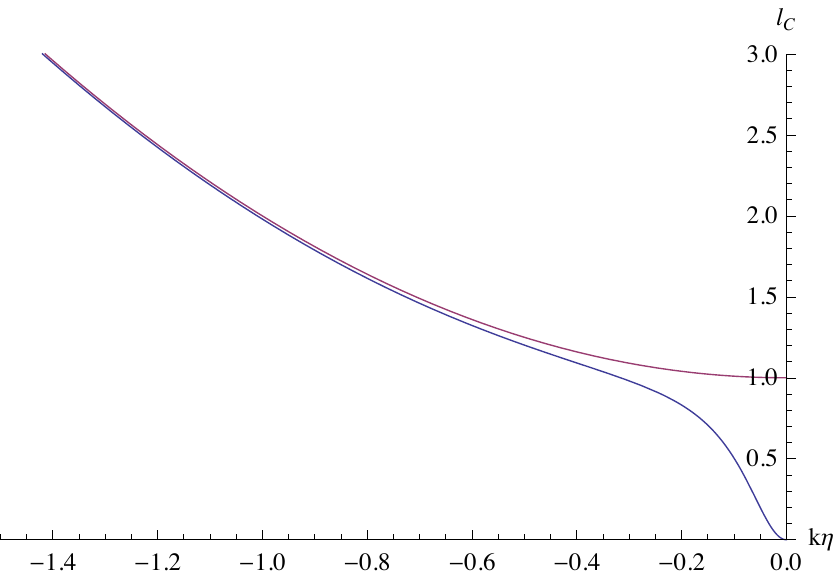}
\caption{The relation of coherence length and the conformal time for usual inflation. The horizontal axis is $k\eta$ and the vertical axis is normalized coherence length. The upper (red) line corresponds to no interaction, and the lower (blue) line corresponds to $c=0.15$. If there is an interaction, the coherence length starts decreasing and eventually becomes zero for the superhorizon modes.}
\label{usualinflation}
\end{figure}

For usual inflation, $a(t)=e^{Ht}$ which is equivalent to $a(\eta)=-\dfrac{1}{H\eta}$. Here, $H$ is the Hubble constant. Eq.~\eqref{uk} then tells us
\begin{equation}
 u_k(\eta)=c_1\frac{e^{-ik\eta}}{\sqrt{2k}}\pbra{1-\frac{i}{k\eta}}+c_2\frac{e^{ik\eta}}{\sqrt{2k}}\pbra{1+\frac{i}{k\eta}}
\end{equation}
Considering the wave functional \eqref{wavefn}, we have to require a positive real part of $A$ for obvious reasons. Therefore, we choose $c_1=0$ and
\begin{equation}
 A_k(\eta)=\frac{k}{H^2\eta^2}\frac{1}{1+\frac{i}{k\eta}}
\end{equation}
Then, Eq.~\eqref{lc} gives us the coherence length \footnote{We recover the results in \cite{Brandenberger:1990bx} after accounting for some typos in that paper.}:
\begin{equation}
 l_c=\frac{H(1+k^2\eta^2)^{1/2}}{k^{3/2}\pbra{1+\frac{c^2}{k^2\eta^2}}^{1/2}}
\end{equation}
 We see that if no interaction is present $(c=0)$, the coherence length approaches a constant value. Adding even a small interaction will reduce it to zero (See Fig.~\ref{usualinflation}). Besides, the coherence length starts to decrease exponentially when the wavelength crosses the Hubble radius, which justifies our heuristic derivation in cosmological perturbation theory.

\section{Decoherence in power law inflation and ekpyrotic phase}
\begin{figure}[ht]
\includegraphics{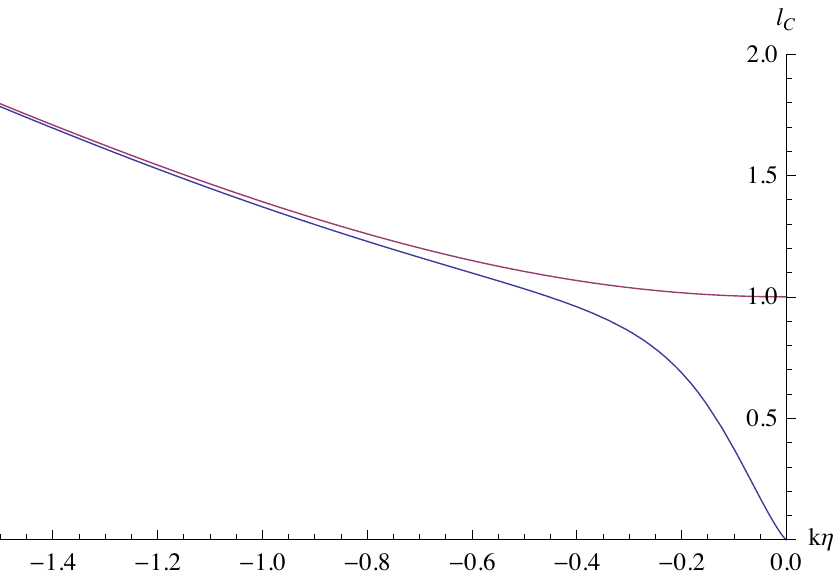} ~~~~~~~~
\caption{The relation of coherence length and the conformal time for power law inflation. We choose $p=10$ in this plot. The upper (red) line corresponds to no interaction, and the lower (blue) line corresponds to $c=0.15$. }
\label{powerlawinflation}
\end{figure}

\begin{figure}[ht]
\includegraphics{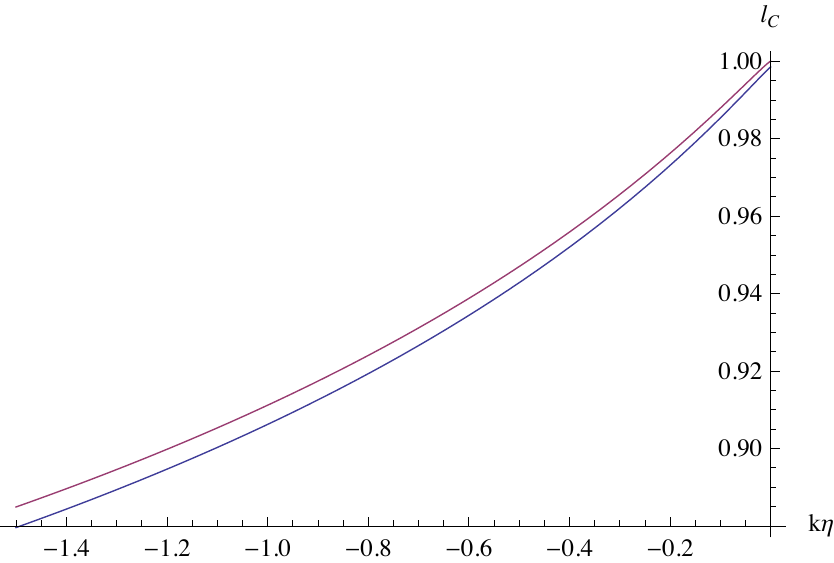} ~~~~~~~~
\caption{The relation of coherence length and the conformal time for ekpyrosis with $p=0.1$. The upper (red) line corresponds to no interaction, and the lower (blue) line corresponds to $c=0.15$. It is clear that even the modes go outside the horizon, the coherence length continues growing and approaches to a nonzero constant in the end.}
\label{ekpyrosis}
\end{figure}

The scale factor behaviors of power law inflation and ekpyrosis are very similar so we consider them at the same time. We list some properties of their scale factors in the Table \ref{compare}.
\begin{center}
\begin{table}
\caption{Table (comparing power law inflation and ekpyrosis)}
  \begin{tabular}{ | c | c | c |}
    \hline
                    & power law inflation      & ekpyrotic phase  \\ \hline \hline
    range of $t$    & $0\leq t\leq \infty$     & $-\infty\leq t\leq 0$ \\ \hline
    $a(t)$          & $t^p$                    & $(-t)^p$              \\ \hline
    $p$             & $p\gg 1$                 & $p\ll1$              \\ \hline
    range of $\eta$ & $-\infty\leq\eta\leq 0$  & $-\infty\leq\eta\leq 0$   \\ \hline
    $a(\eta)$       & $\bbra{(1-p)\eta}^{p/(1-p)}$  & $\bbra{-(1-p)\eta}^{p/(1-p)}$   \\ \hline
    $\frac{\dot a}{a}$    & $ \frac{p}{(1-p)}\frac{1}{\eta}$  & $\frac{p}{(1-p)}\frac{1}{\eta}$   \\ \hline
    $\frac{\ddot a}{a}$    & $ \frac{p(2p-1)}{(1-p)^2}\frac{1}{\eta^2}$  & $\frac{p(2p-1)}{(1-p)^2}\frac{1}{\eta^2}$   \\ \hline
    \hline
  \end{tabular}
\label{compare}
\end{table}
\end{center}
Because both of the power law inflation and ekpyrosis have the same $\dfrac{\ddot a}{a}$, they share the same solution of $u_k$. The differential equation of \eqref{uk} can be solved exactly by
\begin{equation}
 u_k=\sqrt{-k\eta}\bbra{c_1H_\alpha^{(1)}(-k\eta)+c_2H_\alpha^{(2)}(-k\eta)}
\end{equation}
where $H_\alpha^{(1,2)}$ are Hankel functions, and we have defined
\begin{equation}
 \alpha\equiv\sqrt{\frac{\ddot a}{a}\eta^2+\frac 14}=\left|\frac{1-3p}{2(1-p)}\right|
\end{equation}
As before, we want $A_k(\eta)$ to have a positive real part, so we take $c_1=0$, and Eq.~\eqref{Ak} tells us
\begin{equation}
 A_k(\eta)=-ia^2(\eta)\bbra{\frac{1-3p}{2(1-p)}\frac{1}{\eta}-\frac{k}{2}\frac{H_{\alpha-1}^{(2)}(-k\eta)-H_{\alpha+1}^{(2)}(-k\eta)}{H_{\alpha}^{(2)}(-k\eta)}}
\end{equation}
Notice that they are the same for both power law inflation and ekpyrotic phase except $p\gg 1$ for the former and $p\ll 1$ for the latter. We can then use Eq.~\eqref{lc} to calculate the coherence length for both cases. The numerical solutions are plotted in the Fig.~\ref{powerlawinflation} and Fig.~\ref{ekpyrosis}.

In order to get the behavior of the coherence length $l_c$ when the modes are well outside the Hubble radius, we need the asymptotic form of the Hankel function as $x\rightarrow 0$:
\begin{equation}
 H_\alpha^{(2)}(x)\rightarrow\bbra{\frac{1}{\Gamma(\alpha+1)}\pbra{\frac{x}{2}}^\alpha-\frac{1}{\Gamma(\alpha+2)}\pbra{\frac x 2}^{\alpha+2}}+i\bbra{\frac{\Gamma(\alpha)}{\pi}\pbra{\frac x2}^{-\alpha}+\frac{\Gamma(\alpha-1)}{\pi}\pbra{\frac x2}^{2-\alpha}}
\end{equation}
where $\alpha>0$ and $\Gamma(\alpha)$ is the Euler gamma function. After some manipulation of algebra, we have
\begin{eqnarray}
 A_k(\eta) \approx \left\{ \begin{array}{rl}
 2^{1-2\alpha}|1-p|^{1-2\alpha}k^{2\alpha}\bbra{\dfrac{\pi}{\Gamma(\alpha)^2}+i\dfrac{1}{\alpha-1}\pbra{\dfrac{-k\eta}{2}}^{2-2\alpha}} &\text{, if $\alpha>\frac 12$} \\
  2^{1-2\alpha}|1-p|^{1-2\alpha}k^{2\alpha}\bbra{\dfrac{\pi}{\Gamma(\alpha)^2}+i\dfrac{\pi^2}{2\alpha\Gamma(\alpha)^4}\pbra{\dfrac{-k\eta}{2}}^{2\alpha}} &\text{, if $\alpha<\frac 12$}
       \end{array} \right.
\label{Ak_for_ekpyrosis}
\end{eqnarray}
as $-k\eta\ll 1$. 

For power law inflation, $p\gg 1$, we have $\alpha=\frac 32+\frac{1}{p-1}=\frac32+\epsilon$, $0<\epsilon\ll 1$. Therefore, 
\begin{equation}
\label{powerlawlc}
 l_c\approx l_0\bbra{\frac{1}{1+c^2\dfrac{\Gamma(\alpha)^4}{(\alpha-1)^2\pi^2}\pbra{\dfrac{-k\eta}{2}}^{-2-4\epsilon}}}^{\frac 12}
\end{equation}
where
\begin{equation}
 l_0^2=|2(1-p)|^{2+2\epsilon}k^{-3-2\epsilon}\frac{\Gamma(\alpha)^2}{\pi}
\end{equation}
From Eq.~\eqref{powerlawlc}, it is obvious that if no interaction is present, the coherence length approaches a constant value $l_0$. However, even a small interaction will reduce the coherence length to zero just like what happened in the usual inflationary case.

As for the ekpyrotic phase, $p\ll 1$, and $\alpha=\frac12-\frac{p}{1-p}=\frac 12-\epsilon,~0<\epsilon\ll 1$. Use Eq.~\eqref{Ak_for_ekpyrosis}, it is not difficult to get
\begin{equation}
 l_c\approx l_0\bbra{\frac{1}{1+c^2\dfrac{\pi^2}{4\alpha^2\Gamma(\alpha)^4}\pbra{\dfrac{-k\eta}{2}}^{2-4\epsilon}}}^{\frac 12}
\end{equation}
This means the coherence length approaches a nonzero constant value no matter whether the interaction is present or not, in agreement with our numerical results in Fig.~\ref{ekpyrosis}.

\section{Conclusion}
We have studied a simple model with two free scalar fields interacting via a gradient coupling term in three different background spacetime: the usual inflation, the power law inflation, and the ekpyrosis. We also calculate the reduced density matrix and the corresponding coherence length by summing over one of the fields in all three cases. 

Our results are that if no interaction is present, the coherence length approaches a constant value. Adding even a small interation will reduce it to zero in either usual inflation or power law inflation case. Since this decoherence starts at Hubble crossing, the quantum fluctuations evaluated at $k\eta=-1$ give the classical initial density perturbations which become the seeds of inhomogenities of our universe later on. However, this argument does not work for ekpyrosis whose coherence length never hits zero. This means the quantum coherence would not disappear even when the modes leave the horizon. Therefore, the heuristic argument that the quantum fluctuation can become classical for superhorizon modes is not valid for ekpyrotic phase. The implication of our result is that the power spectrum of CMB fluctuations is not directly related to the ekpyrotic phase. Even though at the end of ekpyrosis the scalar field has a scale-invariant power spectrum, it is hard to say anything about what we observe right now, since that depends on the ``classical'' initial density perturbations. This puts some doubts on the analyses of the cosmological perturbations in the cyclic/ekpyrotic universe.  

We derived our results using a very simple model. In principle, if we would like to claim the decoherence phenomenon cannot occur in ekpyrosis, we have to consider all kinds of interactions between systems and environment which is almost impossible to do. However, we believe the basic physics should emerge from  simple models. We can easily generalize our analyses to a massive scalar field, and the results wouldn't change too much. We could also consider different kinds of interactions, but we will leave it to the future work.

Finally, we model the environment with a scalar field, which is convincing but might be an oversimplified assumption. The environment can also be taken to consist of the short wavelength modes which are coupled to the long wavelength modes via non-linear couplings \cite{Calzetta:1995ys,Lombardo:1995fg,Matacz:1996gk,Campo:2005sv,Martineau:2006ki,Burgess:2006jn,Kiefer:2006je}. Hence, this might be another possible way to generate decoherence during ekpyrosis.

\section{Acknowledgements}
This work is supported by the U.S. Department of Energy. We especially thank S.~M.~Carroll for very useful comments and suggestions.

\end{document}